\documentclass[twocolumn,showpacs,amsmath,amssymb,superscriptaddress,aps,pre]{revtex4-1}
\pdfoutput=1
\usepackage{graphicx,color} 
\usepackage{dcolumn}
\usepackage{bm,bbold,empheq,mathdots}
\usepackage{units,enumitem}
\usepackage{hyperref}
\usepackage{eqnarray}

\newcommand{\ba}{\begin{eqnarray}}
\newcommand{\ea}{\end{eqnarray}}
\newcommand{\be}{\begin{equation}}
\newcommand{\ee}{\end{equation}}
\usepackage{amsmath}    
\usepackage{graphicx}   
\usepackage{verbatim}   
\usepackage{color}      
\usepackage{subfigure}  
\usepackage{hyperref}   
\begin{document}

\title{Cross-link induced shrinkage of grafted Gaussian chains}
\author{Panayotis Benetatos}  \email{pben@knu.ac.kr}  \affiliation{Department of Physics, Kyungpook National University, 80 Daehakro, Bukgu, Daegu 702-701, Korea}

\date{\today}  

\begin{abstract}

  The statistical mechanics of polymers grafted on surfaces has been the subject of intense research activity because of many potential applications.  In this paper, we analytically investigate the conformational changes caused by a single cross-link on two ideal (Gaussian) chains grafted on a rigid planar surface. Both the cross-link and the surface reduce the number of allowed configurations. In the absence of the hard substrate, the sole effect of the cross-link is a reduction in the effective Kuhn length of a tethered  chain. The cross-link induced shrinkage (collapse) of the grafted chains (mushrooms) turns out to be a reduction in the variance of the  distribution of the height of the chain rather than a reduction of the height itself.
\end{abstract}


\maketitle

\section{Introduction}

In recent years, there has been a growing interest in polymers grafted (tethered) on surfaces. At high grafting density, polymer brushes allow a tunable modification of the physicochemical properties of the surface with promising technological applications. A means of modifying the properties of a polymer brush is by introducing cross-links between the chains \cite{Ballauff,Loveless,Li}. Grafted polymer structures also exist in biological systems. For example, the thickness elasticity of the red cell membrane has been related to the entropy of spectrin loops grafted on the lipid bilayer \cite{Evans}. Cross-link induced grafted polymer collapse has been proposed as part of the mechanism for the selective gating in the nuclear pore complex which regulates cargo transport between the cytoplasm and the nucleus in eukariotic  cells \cite{Lim,Schoch}.  If the grafting density is low, collective stretching effects become negligible and the conformations of the chains are determined by their bending stiffness and the boundary conditions imposed by the substrate. This is called the mushroom regime. It has been shown that this regime is part of the phase diagram of DNA brushes on a biochip \cite{Bracha}.  In this paper, we focus on the simplest structural element of cross-linked polymer mushrooms which consists of two grafted chains with a single cross-link. Assuming ideal (Gaussian) chains allows exact analytic results.  The conformational probability distribution of a single DNA polymer (mushroom) tethered to a wall has recently been investigated experimentally using the TPM (thethered particle motion) method and good agreement has been found with the Gaussian chain model (for longer strands) \cite{Lindner1,Lindner2}.

The paper is organized as follows: In Section II we introduce the model of two grafted Gaussian chains on a planar substrate with a single cross-link. In Section III, we show that, in the absence of the substrate, the cross-linked chain behaves as a free Gaussian chain with a renormalized persistence length. Using the appropriate propagators,  we calculate the probability distribution of a cross-linked mushroom's free end. In addition to the change in the height distribution, the cross-link also pulls the chains closer to each other thus causing a lateral displacement which is analyzed in Section IV.  We conclude and discuss further extensions of this work in Section V.

\section{Model}

\begin{figure}
\includegraphics[
width=0.40\textwidth]{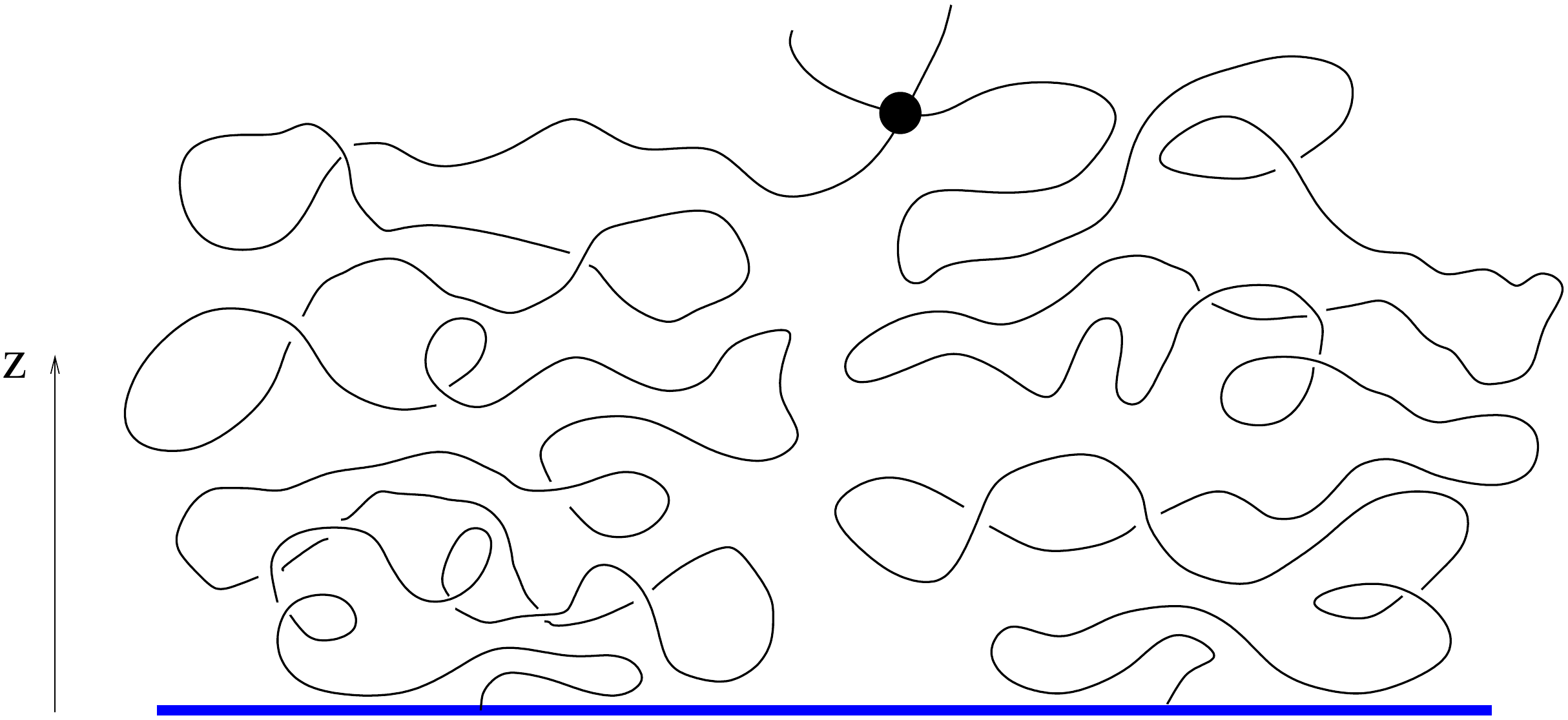}\caption{(Color online) A schematic diagram of two ideal chains (mushrooms) with a single cross-link grafted on a rigid planar substrate.
\label{schematic}}
\end{figure}

We consider two identical flexible polymers, modeled as Gaussian chains, grafted at different points on a flat substrate with absorbing boundary conditions (i.e., the polymers are repelled from the grafting surface by a hard-wall potential) as shown in Fig. \ref{schematic}. The substrate is the $xy$-plane whereas the polymers are confined in the upper half-space.  In the absence of the substrate, and without any cross-link, the conformational probability of each chain would be given by the Wiener measure \cite{Doi-Edwards}:
\begin{equation}
P(\{{\bf r}_i(s)\})\sim \exp\Big[-\frac{3}{4l_p}\int_0^L\Big(\frac{\partial {\bf r}_i}{\partial s}\Big)^2 ds\Big]\;,
\end{equation}
where $l_p$ is the microscopic persistence length (one half of the Kuhn length) and $L$ is the contour length. The grafted ends are: ${\bf r}_1(0)=(0,0,0)={\bf r}_{10}$ and ${\bf r}_2(0)=(d,0,0)={\bf r}_{20}$. 

A single cross-link is modeled as a harmonic spring connecting ${\bf r}_1(s_{c1})$ to ${\bf r}_2(s_{c2})$. The Edwards Hamiltonian (effective free energy functional) of the cross-linked system of the two chains reads:
\begin{align}
\label{Hamiltonian1}
& {\cal H}(\{{\bf r}_1(s)\}, \{{\bf r}_2(s)\}) =\frac{3}{4l_p\beta}\int_0^L\Big(\frac{\partial {\bf r}_1}{\partial s}\Big)^2 ds \\ \nonumber
 &+ \frac{3}{4l_p\beta}\int_0^L\Big(\frac{\partial {\bf r}_2}{\partial s}\Big)^2 ds + \frac{g}{2}\big({\bf r}_1(s_{c1})-{\bf r}_2(s_{c2})\big)^2\\ \nonumber
 &+\int_0^L\Big(U\big({\bf r}_1(s)\big)+U\big({\bf r}_2(s)\big)\Big)ds\;,
 \end{align}
where $\beta=1/k_BT$, and the cross-link strength $g$ is related to a temperature-independent cross-link size $a_c$ via $\beta g =3/(a_c^2)$. The presence of the substrate is expressed by the hrad-wall potential $U(x,y,z)=0$  for $z>0$ and $U(x,y,z)=+\infty$ for $z<0$. The joint probability distribution for the end points (propagator or Green's function) is the path integral:
\begin{equation}
G({\bf r}_1,{\bf r}_2 |{\bf r}_{10},{\bf r}_{20})=\int_{{\bf r}_1(0)={\bf r}_{10}}^{{\bf r}_1(L)={\bf r}_{1}} \int_{{\bf r}_2(0)={\bf r}_{20}}^{{\bf r}_2(L)={\bf r}_{2}}  e^{-\beta {\cal H}(\{{\bf r}_1(s)\}, \{{\bf r}_2(s)\})} \;.
\label{propagator}
\end{equation}
The presence of the substrate imposes the half-space constraint which is encoded in the absorbing boundary condition \cite{Chandra,Dimargio}:
\begin{equation}
G({\bf r}_1(s),{\bf r}_2 (s) | {\bf r}_{10},{\bf r}_{20})=0,\;\;\; \forall \;\;z_i (s) = 0 \;.
\label{absorbing}
\end{equation}

The three directions $(x, y, z)$ decouple in the Hamiltonian of Eq. \eqref{Hamiltonian1} and the propagator factorizes accordingly. Since we are interested primarily in the height of the cross-linked polymers, we focus on the one-dimensional problem characterized by the reduced Hamiltonian
\begin{align}
\label{Hamiltonian2}
& {\cal H}_z(\{z_1(s)\}, \{z_2(s)\}) =\frac{3}{4l_p\beta}\int_0^L\Big(\frac{\partial z_1}{\partial s}\Big)^2 ds \\ \nonumber
 &+ \frac{3}{4l_p\beta}\int_0^L\Big(\frac{\partial z_2}{\partial s}\Big)^2 ds + \frac{g}{2}\big(z_1(s_{c1})-z_2(s_{c2})\big)^2\\ \nonumber
&+ \int_0^L\Big(U\big(z_1(s)\big)+U\big(z_2(s)\big)\Big)ds\;.
\end{align}

\section{Height of the cross-linked mushrooms}

Our goal is to calculate the effect of the cross-link expressed by the third term in the Hamiltionian \eqref{Hamiltonian2} on the height of mushroom $1$ expressed by the average distance $z_1(L)$. We first notice that, in the long-chain limit and in the absence of the substrate, integrating out the degrees of freedom of the second chain, yields a Gaussian chain with a renormalized persistence length $\tilde{l}_p$. Thus our first step will be to calculate the effect of the cross-link in the absence of the absorbing boundary conditions imposed by the substrate. 

We expand $z_1(s)$ and $z_2(s)$ in the appropriate Fourier modes \\
\begin{align}
& z_1(s)=\sum\limits_{n=0}^{\infty}A_n \sin(q_n s) \\ \nonumber
& z_2(s)=\sum\limits_{n=0}^{\infty}B_n \sin(q_n s) \;,
\end{align}
where $q_n=\frac{\pi}{2L}(2n+1)$ with $n=0,1,2,..$, in accordance with  the grafted-free boundary conditions $z_i(s=0)=0$ and $\frac{dz_i}{ds}(s=L)=0$. With this substitution, the Hamiltonian \eqref{Hamiltonian2} (without the potential $U$) is written as
\begin{align}
\label{Hamiltonian3}
 &{\cal H}_z(A_n, B_n) =\frac{3L}{4l_p\beta}\sum\limits_{n=0}^{\infty}\frac{q_n^2}{2}A_n^2\\ \nonumber
 +& \frac{3L}{4l_p\beta}\sum\limits_{n=0}^{\infty}\frac{q_n^2}{2}B_n^2\\ \nonumber
 + \frac{g}{2}\Big\{&\sum\limits_{n=0}^{\infty}\sum\limits_{m=0}^{\infty}A_n A_m \sin(q_n s_{c1})\sin(q_m s_{c1})\\ \nonumber
 +&\sum\limits_{n=0}^{\infty}\sum\limits_{m=0}^{\infty}B_n B_m \sin(q_n s_{c2})\sin(q_m s_{c2})\\ \nonumber
 -&2\sum\limits_{n=0}^{\infty}\sum\limits_{m=0}^{\infty}A_n B_m \sin(q_n s_{c1})\sin(q_m s_{c2})\Big\}\;.
\end{align}
Introducing the culumn vector $\Gamma$ such that 
\begin{align}
\Gamma^T=(A_0,B_0,A_1,B_1,...)
\label{Gamma}
\end{align}
and the matrix $C$ such that
\begin{gather*}
C=\frac{3L}{4l_p\beta}
\begin{pmatrix}
q_0^2 & 0 & 0 & 0 & \hdots \\
0 & q_0^2 & 0 &  0 &\hdots \\
0 & 0 & q_1^2 & 0 &  \hdots \\
0 & 0 &  0& q_1^2 & \hdots \\
\vdots & \vdots & \vdots & \vdots & \ddots
\end{pmatrix}\;,
\end{gather*}
the Hamiltionian is rewritten in a compact form as
\begin{align}
\label{Hamiltonian4}
H_z(\Gamma)=\frac{1}{2}\sum\limits_{n=0}^{\infty}\sum\limits_{m=0}^{\infty}\Gamma_n G_{nm} \Gamma_m\;,
\end{align}
where $G_{nm}=C_{nm}+u_n u_m$ with the column vector $u$ defined such that
\begin{align}
\label{u}
u^T=\sqrt{g}\big(\sin(q_0s_{c1}), -\sin(q_0s_{c2}),\sin(q_1s_{c1}),-\sin(q_2s_{c2}),...\big)
\end{align}

 In order to calculate the mean square height of the free end (where $s=L$) of chain $1$, we need to invert matrix G. This is easily done using the Sherman-Morrison formula from linear algebra \cite{Teukolsky}. We obtain:
 \begin{align}
\label{Sherman_Morrison}
 G^{-1}=C^{-1}-\frac{C^{-1}uu^TC^{-1}}{1+u^TC^{-1}u}\;.
\end{align}  
 The mean square height of chain $1$ is given by\\
 \begin{eqnarray}
\label{height}
  \langle z_1^2(L)\rangle&=&\sum\limits_{n=0}^{\infty}\sum\limits_{m=0}^{\infty}\langle A_n A_m \rangle \sin(q_n L)\sin(q_m L)\\ \nonumber
 &=&\sum\limits_{n=0}^{\infty}\sum\limits_{m=0}^{\infty}(G^{-1})_{2n,2m}(-1)^n(-1)^m\;.
\end{eqnarray}    
The first term in the rhs of Eq. \eqref{Sherman_Morrison} contributes $\sum\limits_{n=0}^{\infty}\sum\limits_{m=0}^{\infty}\frac{4l_p}{3L}q_n^{-2}\delta_{nm}(-1)^n(-1)^m=(2/3)l_pL$ as expected from an uncross-linked Gaussian chain. The effect of the cross-link is expressed by the second term. One can easily see that the term $u^TC^{-1}u$ in the denominator scales as $\sim L$ and the numerator, $C^{-1}uu^TC^{-1}$, scales as $\sim L^2$. Therefore, in the long-chain limit, this term scales as $\sim L$. The prefactor involves a fraction of $l_p$ which can be interpreted as the cross-link induced renormalization of $l_p$ for chain 1. Its calculation entails the summation of two series which can easily be done, for arbitrary $s_{c1}$ and $s_{c2}$ numerically. For $s_{c1}=L\;{\rm or}\;L/2$ and $s_{c2}=L\;{\rm or}\;L/2$ the summation can be done analytically. 
     
    For a cross-link connecting the midpoints of the two polymers, $s_{c1}=s_{c2}=L/2$ and Eqs. \eqref{Sherman_Morrison} and \eqref{height} give:
  \begin{align}
  \label{scL/2}
  \langle z_1^2(L) \rangle =\frac{2}{3}l_p L - \frac{\beta g}{4}\frac{\big(\frac{2}{3}l_pL\big)^2}{1+\beta g\frac{2}{3}l_pL}
  \xrightarrow{L\gg l_p} \frac{2}{3}\frac{3 l_p}{4} L\;.
      \end{align}
   We notice that the hard cross-link limit ($g\rightarrow\infty$) is equivalent to the long chain ($L\gg l_p$) limit, as expected. The effect of the cross-link is a reduction in the height of the polymer which can be expressed as an effective reduction of the persistence length by one quarter. For a cross-link connecting the endpoints of the two polymers, $s_{c1}=s_{c2}=L$ and Eqs. \eqref{Sherman_Morrison} and \eqref{height} give:
  \begin{align}
  \label{scL}
  \langle z_1^2(L) \rangle =\frac{2}{3}l_p L - \beta g\frac{\big(\frac{2}{3}l_pL\big)^2}{1+2\beta g\frac{2}{3}l_pL}
  \xrightarrow{L\gg l_p} \frac{2}{3}\frac{ l_p}{2} L\;.
      \end{align}
  The effect of this end-link is a reduction in the height of the polymer which can be expressed as  an effective reduction of its persistence length by half.
     
  A more efficient way to obtain the previous results is to use the method of propagators (Green's functions). The probability distribution of finding monomer $s'$ of a free Gaussian chain at height $z(s')$, given that monomer $s$ is at height $z(s)$ ($s'>s$), is given by the propagator:
  \begin{align}
 \label{propagator}
   K[z(s'),z(s)]={\cal N} \exp\Big(-\frac{3 \big(z(s')-z(s)\big)^2}{4 l_p (s'-s)}\Big)\;,
  \end{align}
  where ${\cal N}$ is a normalization constant. The probability distribution of the free end of chain $1$ which is cross-linked with chain $2$ according to \eqref{Hamiltonian2} is given by
\begin{widetext}
\begin{align}
P[z_1(L)]&\sim \int_{-\infty}^{\infty}\int_{-\infty}^{\infty}d[z_1(s_{c1})]d[z_2(s_{c2})]K[z_1(s_{c1}),0(0)]K[z_2(s_{c2}),0(0)]\exp\Big(-\frac{\beta g}{2} \big(z_1(s_{c1})-z_2(s_{c2})\big)^2\Big)K[z_1(L),z_1(s_{c1})]\\ \nonumber
&\sim\exp\Big(-\frac{\frac{2 \beta g}{3}(s_{c1}+s_{c2})l_p+1}{\frac{2 \beta g l_p}{3}(s_{c1}L+s_{c2}L-s_{c1}^2)+L}\frac{3 z_1^2(L)}{4 l_p} \Big)\xrightarrow{g\rightarrow \infty}\exp\Big(-\frac{s_{c1}+s_{c2}}{s_{c1}L+s_{c2}L-s_{c1}^2}\frac{3 z_1^2(L)}{4 l_p} \Big)
 \end{align}
 \end{widetext}   
 We point out that this distribution exaclty reproduces our previous results for $\langle z_1^2(L)\rangle$.

 In order to take into account the effect of the substrate, we need the propagator of a free Gaussian chain adjusted by the absorbing boundary condition. As shown in \cite{Dolan_Edwards} and \cite{Slutsky}, this problem can be solved by the method of images. Just as in electrostatics, we place a virtual source similar to an image charge of the opposite sign on the opposite side of the plane in such a way as to enforce the required boundary condition:
 \begin{align}
 &\tilde{K}[z(s'),z(s)]=\\ \nonumber
 &\frac{\exp\Big(-\frac{3 \big(z(s')-z(s)\big)^2}{4 l_p (s'-s)}\Big)-\exp\Big(-\frac{3 \big(z(s')+z(s)\big)^2}{4 l_p (s'-s)}\Big)}{\sqrt{4\pi l_p (s'-s)/3}}
\end{align}
 If the starting point lies on the substrate ($z(s)=0$), for regularization reasons which are explained in \cite{Slutsky} and \cite{Deam_Edwards}, we shift it slightly above it by a microscopic length $a\approx l_p$ thus getting:
  \begin{align}
  \tilde{K}[z(s'),a(s)]\sim\frac{3a(s) z(s')}{l_p (s'-s)}\exp\Big(-\frac{3 z^2(s')}{4 l_p(s'-s)}\Big)
  \end{align} 
 which is a Rayleigh function of $z(s')$.   
      
Using the abovementioned  propagators,  the probability distribution of the free end of the cross-linked chain $1$ reads
\begin{widetext}
\begin{align}
P[z_1(L)]&\sim \int_0^{\infty}\int_0^{\infty}d[z_1(s_{c1})]d[z_2(s_{c2})]\tilde{K}[z_1(s_{c1}),a(0)]\tilde{K}[z_2(s_{c2}),a(0)]\exp\Big(-\frac{\beta g}{2} \big(z_1(s_{c1})-z_2(s_{c2})\big)^2\Big)\tilde{K}[z_1(L),z_1(s_{c1})]\\ \nonumber
&\xrightarrow{g\rightarrow\infty} \;\sim6\sqrt{s_{c1}s_{c2}l_p(L(s_{c1}+s_{c2})-s_{c1}^2)}(L-s_{c1})z_1(L)\exp\Big(-\frac{3z_1^2(L)}{4l_p(L-s_{c1})}\Big)+\sqrt{3\pi (L-s_{c1})}(2s_{c1}l_pL^2+2s_{c2}l_pL^2\\ \nonumber
&-4s_{c1}^2l_pL-2s_{c2}s_{c1}l_pL+3s_{c2}s_{c1}z_1^2(L)+2s_{c1}^3l_p)\exp\Big(-\frac{3(s_{c2}+s_{c1})z_1^2(L)}{l_p(L(s_{c1}+s_{c2})-s_{c1}^2)}\Big)\\ \nonumber
&\times {\rm erf}\Big(\frac{\sqrt{3 s_{c1}s_{c2}} z_1(L)}{2\sqrt{l_p(L-s_{c1})(L(s_{c1}+s_{c2})-s_{c1}^2)}}\Big)\;.
 \end{align}
 \end{widetext}                                                                                  
 We notice that this distribution has a profile qualitatively similar to that of the Rayleigh function but it is not a Rayleigh function. In order to gain some insight into the combined effect of the cross-link and the substrate, let us consider the case of a single hard end-link:                                                                                                                   $s_{c1}=s_{c2}=L$. The above formula gives: $P[z_1(L)]\sim z_1^2(L)\exp(-3z_1^2(L)/(2l_pL))$. The exponential indicates a reduction of the effective persistence length by half in accord to what we get in the absence of the substrate. But the prefactor no longer is  linear and has changed to quadratic. Thus we conclude that cross-linking in the presence of a planar hard wall is not simply equivalent to a renormalization of the persistence length. We can generalize this result to $n$ identical end-linked mushrooms. In that case, we would get: $P[z_1(L)]\sim z_1^n(L)\exp(-3 n z_1^2(L)/(4l_pL))$. It is interesting to notice that the resulting distribution of $z_1(L)$ maintains its maximum at the same height irrespective of the number of end-linked chains. What changes is the variance (spread) of the distribution about its peak, which decreases with $n$.
                                                                                                                                                  
     In the Gaussian chain approximation which is used in the above analysis, the effect of the cross-link is independent of the distance $d$ between the two grafted points. For this approximation to hold, the lateral stretching due to the cross-link must be small compared to the total stored length: $d\ll s_{c1}+s_{c2}$ (for a hard cross-link).

     \section{Lateral Displacement}
   Because the two chains are Gaussian and the three dimensions in the Hamiltonian decouple, the substrate will not affect the lateral displacement. Let us define a Cartesian coordinate system where the grafting point of chain $1$ is at the origin and the $x$-axis is in the line connecting the grafting points of the two chains. The relevant part of the Hamiltonian reads
 \begin{align}
\label{Hamiltonianx}
& {\cal H}_x(\{z_1(s)\}, \{x_2(s)\}) =\frac{3}{4l_p\beta}\int_0^L\Big(\frac{\partial x_1}{\partial s}\Big)^2 ds \\ \nonumber
 &+ \frac{3}{4l_p\beta}\int_0^L\Big(\frac{\partial x_2}{\partial s}\Big)^2 ds + \frac{g}{2}\big(x_1(s_{c1})-x_2(s_{c2})\big)^2\;.
\end{align}  
   We expand in the appropriate Fourier modes:
 \begin{align}
& x_1(s)=\sum\limits_{n=0}^{\infty}A_n \sin(q_n s) \\ \nonumber
& x_2(s)=d+\sum\limits_{n=0}^{\infty}B_n \sin(q_n s) \;,
\end{align}
where  $d$ is the distance between the two grafting points and the wavenumbers $q_n$ are defined as in Section III. Keeping the notation of the previous section, ${\cal H}_x$ can be written in compact for as
\begin{align}
\label{Hamiltonianx1}
 H_x(\Gamma)=\frac{1}{2}\sum\limits_{n=0}^{\infty}\sum\limits_{m=0}^{\infty}\Gamma_n G_{nm} \Gamma_m+\sum\limits_{n=0}^{\infty}D_n\Gamma_n\;,
\end{align}                   
where the vector $D$ is defined as $D=-\sqrt{g}du$, and we have omitted an irrelevant constant term $gd^2/2$.                                                            
                                   
 In order to calculate $\langle x_1(L)\rangle$, we use 
 \begin{align}                              
  \langle \Gamma_n \rangle = -\sum\limits_{m=0}^{\infty}    (G^{-1})_{nm} D_m                             
  \end{align}                                 
 The series involved in the calculation, can easily be summed numerically for arbitrary                                 link points. In the following, we present analytic results for three special cases.
                                   
 For $s_{c1}=s_{c2}=L$, we obtain
 \begin{align}
 \langle x_1(L)\rangle=\beta gd\frac{2l_pL}{3}-\frac{(\beta gd2l_pL/3)(4\beta gl_pL/3)}{1+4\beta gl_pL/3}\xrightarrow{g\rightarrow \infty}\frac{d}{2}\;.
 \end{align}                           
 The hard cross-link limit ($g\rightarrow \infty$) yields the result that we expect from the symmetry of the system.  For $s_{c1}=s_c{2}=L/2$, we obtain
 \begin{align}
  \langle x_1(L)\rangle=\beta gd\frac{l_pL}{3}-\frac{(\beta gdl_pL/3)(2\beta gl_pL/3)}{1+2\beta gl_pL/3} \xrightarrow{g\rightarrow \infty}\frac{d}{2}\;.
\end{align}
 Again, the hard cross-link limit yields the result which is expected from the symmetry. For $s_{c1}=L/2$ and $s_{c2}=L$, we obtain
\begin{align}
  \langle x_1(L)\rangle=\beta gd\frac{l_pL}{3}-\frac{(\beta gdl_pL/3)(\beta gl_pL)}{1+\beta gl_pL} 
 \xrightarrow{g\rightarrow \infty}\frac{d}{3}\;. 
 \end{align} 

  As in Section II, the method of propagators turns out to be more efficient. Because of the decoupling of the three dimensions in the ideal chain, the appropriate propagator is the one given in \eqref{propagator}. Thus we obtain the probability distribution                         for $x_1(L)$:
  \begin{widetext}
\begin{align}
P[x_1(L)]&\sim \int_{-\infty}^{\infty}\int_{-\infty}^{\infty}d[x_1(s_{c1})]d[x_2(s_{c2})]K[x_1(s_{c1}),0(0)]K[x_2(s_{c2}),d(0)]\exp\Big(-\frac{\beta g}{2} \big(x_1(s_{c1})-x_2(s_{c2})\big)^2\Big)K[x_1(L),x_1(s_{c1})]\\ \nonumber
&\sim \exp\Big(-\frac{3}{4 l_p}\frac{4L(\beta g/2)d^2 l_p - 8 (\beta g / 2)d l_p s_{c1} x_1(L)+(4(\beta g/2)l_p( s_{c1}+s_{c2})+3)x_1^2(L)}{4l_p L (\beta g/2)(s_{c1}+s_{c2})-4lp (\beta g/2)s_{c1}^2+3L}\Big)\\ \nonumber
&\xrightarrow{g\rightarrow \infty}\exp\Big(-\frac{3}{4}\frac{L d^2 - 2d s_{c1} x_1(L)+x_1^2(L)(s_{c1}+s_{c2})}{l_p L(s_{sc1}+s_{c2})-l_ps_{c1}^2}\Big)\;.
  \end{align}
  \end{widetext}                                                                  
 From the above distribution, for a hard cross-link, we obtain
 \begin{align}
 \langle x_1(L) \rangle = \frac{d s_{c1}}{s_{c1}+s_{c2}}
 \end{align}
 which agrees with the previous results.

                                   \section{Discussion - Conclusions}
             
             In this paper, we investigated the effect of a single cross-link on two tethered Gaussian chains. In the absence of a hard-wall substrate, the free end of a cross-linked chain behaves as a Gaussian chain with a decreased persistence length. We obtained this result using two methods: explicit calulation of the functional integral and the propagator (Green's function) method. We calculated the probability distribution of the free end of a cross-linked ideal mushroom using the appropriate propagator. We notice that, in the presence of the substrate, the free end no longer behaves as random walk in half-space (Rayleigh function). The main effect of the cross-link is to reduce the variance of the distribution. 
             
                           This work could be generalized to take into account more cross-links, more chains, or more complicated and experimentally relevant geometries for the grafting surface \cite{Kantor}. The role of permanent cross-links on brushes of randomly grafted chains is an interesting open question. On the one hand, cross-links on freely sliding polymers tend to lead to the formation of an amorphous  gel \cite{PBdps}. On the other hand, cross-links give rise to an effective attraction and it has been shown that brushes with attractive interactions exhibit a bundling instability \cite{PBbundle}. Which transition prevails remains to be explored.

\begin{acknowledgements}

I acknowledge support by Kyungpook National University Research Fund, 2013(2014). I also thank the COSA Group at the NCSR Demokritos in Athens, Greece, for hospitality during part of this work.
\end{acknowledgements}

\end{document}